\documentclass[10pt, conference]{IEEEtran}
\usepackage{cite}
\usepackage{amsmath,amssymb,amsfonts}
\usepackage{algorithmic}
\usepackage{graphicx}
\usepackage{textcomp}
\usepackage{xcolor}
\usepackage{hyperref}
\hypersetup{draft}
\def\BibTeX{{\rm B\kern-.05em{\sc i\kern-.025em b}\kern-.08em
    T\kern-.1667em\lower.7ex\hbox{E}\kern-.125emX}}
    
\begin{document}

\title{\textsc{Heimdall}: an AI-based infrastructure for traffic monitoring and anomalies detection}

\author{\IEEEauthorblockN{Andrea Atzori\IEEEauthorrefmark{1}, Silvio Barra\IEEEauthorrefmark{2}, Salvatore Carta\IEEEauthorrefmark{1}, Gianni Fenu\IEEEauthorrefmark{1} and Alessandro Sebastian Podda\IEEEauthorrefmark{1}}
\vspace{1em}
\IEEEauthorblockA{\IEEEauthorrefmark{1}\textit{Department of Mathematics and Computer Science, University of Cagliari}\\
Via Ospedale 72, 09124 Cagliari (Italy)\\
Email: a.atzori28@studenti.unica.it, \{salvatore, fenu, sebastianpodda\}@unica.it}
\vspace{1em}
\IEEEauthorblockA{\IEEEauthorrefmark{2}\textit{Dept. of Electric Engineering and Information Technology, University of Naples ``Federico II"}\\
Via Claudio 21, 80125 Napoli (Italy)\\
Email: silvio.barra@unina.it}}

\maketitle

\begin{abstract}
Since their appearance, Smart Cities have aimed at improving the daily life of people, helping to make public services smarter and more efficient. Several of these services are often intended to provide better security conditions for citizens and drivers. In this vein, we present \textsc{Heimdall}, an AI-based video surveillance system for traffic monitoring and anomalies detection. The proposed system features three main tiers: a ground level, consisting of a set of smart lampposts equipped with cameras and sensors, and an advanced AI unit for detecting accidents and traffic anomalies in real time; a territorial level, which integrates and combines the information collected from the different lampposts, and cross-correlates it with external data sources, in order to coordinate and handle warnings and alerts; a training level, in charge of continuously improving the accuracy of the modules that have to sense the environment. Finally, we propose and discuss an early experimental approach for the detection of anomalies, based on a Faster R-CNN, and adopted in the proposed infrastructure. 
\end{abstract}


\section{Introduction} \label{intro}

The times in which we live are characterised by an ever-increasing adoption of the so-called \emph{smart objects}. This definition arises from the fact that the use of such objects is able to confer ``smartness" to a specific environment. Indeed, these devices are actually sensors, in charge of three main goals:
\begin{itemize}
    \item \textit{sense} the environment;
    \item \textit{collect} information from the environment;
    \item \textit{take actions} within the environment;
\end{itemize}
and their application to the wide range contexts of everyday life has given rise to concepts such as those of Smart City, Smart Home, Smart Car, to name a few.

Nowadays, scientific and industrial research are very active in this direction, since it involves a large number of topics, including, e.g., pattern recognition applications, cloud-based systems, cryptography and several others. In this regard, Smart Cities  certainly represent one of the most studied environments, since the definition of Internet of Things (IoT) takes here a wider and more interesting meaning. 
In fact, many applications take advantages from the adoption of IoT networks in Smart Cities, including (but not limited to) biometrics\cite{barra2018cloud,de2019walking}, traffic light management\cite{liang2019deep,you2020traffic}, video surveillance\cite{rego2018intelligent,neves2015acquiring}, crowd analysis \cite{liu2019context} (especially in situations in which gathering of people is not allowed \cite{chick2020using}) and anomalies detection.

\begin{figure*}[htp!]
\vspace{0.2em}
\centering
\includegraphics[width=0.98\linewidth]{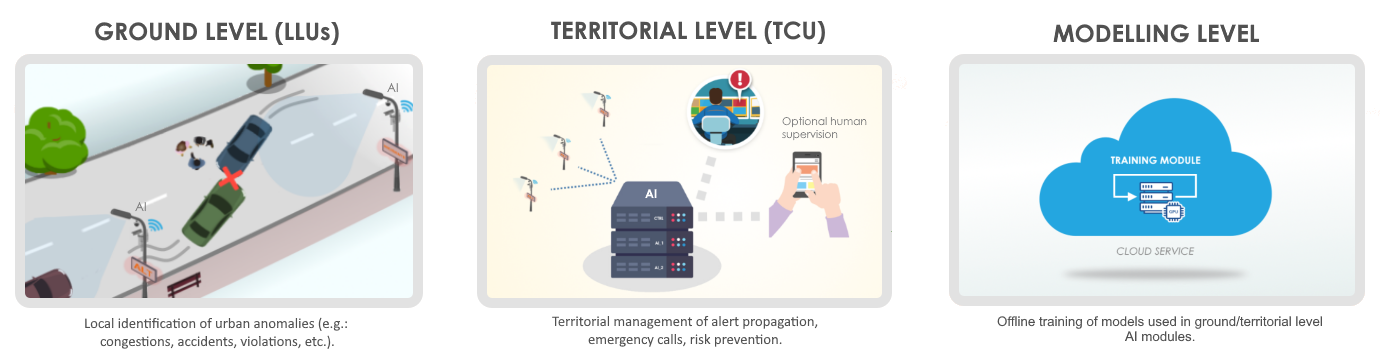}
\caption{The three main subsystems of Heimdall.}
\label{fig:architecture}
\end{figure*}

However, given the \textit{in-the-wild} nature of the urban scenarios, many of the aforementioned applications need to be tailored to the difficulties and challenges of such a typical unconstrained environment: here, visual sensors like cameras may present issues due to light variations, pose, occlusion and perspective \cite{neves2015quis}; microphones may suffer interference from nearby devices or distortions when loud noises are recorded \cite{van2011ability}; wearable sensors are often affected by the sudden movements of the subject which may tend to cause artifact and noise during the acquisition stage \cite{abate2018you}.

Additional difficulties may also arise from the management of master-slave camera systems, where the direction, coordination, calibration and communication of cameras is fundamental for a proper video surveillance activity \cite{neves2015calibration}. Nevertheless, data heterogeneity is another issue which designers need to deal with. The challenge, from this perspective, lies in the fact that different sources gather different data in different format. Fusion algorithms may therefore operate at different levels (sensor, feature vector, score and decision), hence resulting in different grades of precision \cite{barra2014complex}.

In the light of the above, in this paper we present \textsc{Heimdall}, an AI-based video surveillance infrastructure for traffic monitoring in Smart Cities. The main contributions of this paper are summarized below:
\begin{itemize}
    \item the design and description of an end-to-end IoT system for detecting and preventing traffic anomalies that may potentially cause danger to road users, within a urban scenario;
    \item a novel three-tier architecture, in which a ground layer consisting of a set of autonomous smart lamppost is coordinated by an advanced centralised unit that acts at a territorial level, both supported by a modelling layer that continuously trains and improve the involved AI modules;
    \item an early experimental approach for the automatic detection of anomalies, that leverages on Deep Neural Networks (DNNs), and in which the training is carried out by combining synthetic and real-world datasets in order to achieve high effectiveness in detecting both static and dynamic anomalies;
    \item last generation NVIDIA Jetson AGX Xavier card is used as integrated hardware board embedded in the lampost.
\end{itemize}

The remaining of the paper is organized as it follows: Section~\ref{sec:related} briefly describes the related work in the literature. Section~\ref{sec:architecture} presents the architecture of \textsc{Heimdall}, while Section~\ref{sec:anomalies} illustrates the preliminary proposed approach for the automatic detection of anomalies. Finally, in Section~\ref{sec:conclusions}, a timely discussion of the main problems and limitations related to the used techniques is inserted, with a particular emphasis on the proposed infrastructure.

\section{State of the art of the current systems} \label{sec:related}

The massive deployment of the video surveillance infrastructures in modern cities is mainly due to the request for security and safety for pedestrians and drivers \cite{9171212,edinger2019wheelshare}. Systems aimed at ensuring road safety have been active in various parts of the world between the second half of the 90s and the beginning of the new century \cite{regazzoni1998advanced,collins2000introduction}, when video cameras were installed at road junctions in many metropolitan areas, to monitor road sections. In most cases, accident and anomalies detection were carried out using feature extraction techniques that based their operation on the study and generalisation of patterns within images and videos, to recover the same pattern from successive video streams, in times far removed from real time. Many of these techniques are still used today \cite{kamijo2000traffic}, but have given way to methodologies that allow for a better generalization and exploit more advanced tools. Indeed, in the last decade, the theoretical background on artificial intelligence, introduced and deepened in the 80s, have began to show the first practical applications, thanks to the adoption of new high-performance processing systems, thus leading to the fast emergence of modern Machine Learning techniques and, subsequently, of their more extensive branch, the Deep Learning. These methods have undermined canonical feature extraction techniques, which, although they do not require much data to work acceptably, do not always guarantee high reliability and accuracy at the generalisation stage. Deep Learning is now widely used in video and image analysis in a variety of research areas, ranging from medical diagnostics \cite{bibault2020deep, swid2020detection}, financial forecasting \cite{barra2020deep, carta2020multilayer} and cybersecurity \cite{ismail2020deep}, and over the years has also had a positive impact on road safety \cite{parsa2020toward}. The state of the art presents considerable scientific solutions to the problem of road safety, providing inspiration for the development of methods for pedestrian detection \cite{helali2020hardware}, vehicle tracking \cite{do2019visible}, and accident detection \cite{samia2020review}. 
The solution proposed for this project, compared to the state-of-the-art, presents several interesting innovations, introduced with the aim of creating a safe and safeguarded environment for vehicles and pedestrians. First of all, its architecture composed of environmental sensors, hence strongly based on the Internet Of Things concept, brings the idea into the Smart Cities domain, within which so far only theoretical frameworks, with a limited development phase, have been proposed \cite{davydov2020accident}. On the other hand, from a technical point of view, the proposed infrastructure employs video-summarization methodologies for the reduction of the computational load to be transmitted to central and geographically adjacent processing systems. This issue is very much in vogue nowadays, especially with regard to Edge Computing \cite{barra2019biometric} and thus new research directions that aim to migrate part of the processing effort to systems on the edge of the cloud. 
\section{Heimdall: the Al(S)ighty}
\label{sec:systemdescription}
\label{sec:architecture}

The goal of \textsc{Heimdall} is to automatically detect traffic anomalies in urban or extra-urban scenarios, mainly in order to prevent critical situations and accidents. In this context, the system exploits the state of the art of AI technologies (specifically, those based on Machine Learning and Deep Learning techniques), in order to support the action of traffic wardens, as well as to implement appropriate warning measures to protect vehicles and pedestrians in the involved areas.

\textsc{Heimdall} consists of three main subsystems, strongly interconnected, whose coordinated action guarantees the effectiveness of detection and prevention. These subsystems, depicted in Figure \ref{fig:architecture}, can be summarized as it follows: 
\begin{itemize}
    \item the \textit{ground level}, composed by a large number of local AI-based units, embedded on roadside smart lampposts, for real-time detection of traffic anomalies and accidents;
    \item the \textit{territorial level}, which collects data from the lampposts, combining them with information from external sources (e.g. weather, civil protection, traffic conditions), in order to globally manage the alerts, propagate them and assist the human operator responsible for supervising the territory;
    \item the \textit{modelling level}, that consists of an offline training module to upgrade the neural networks according to the newly observed traffic data.
\end{itemize}

In next subsections, we now provide additional details on the aforementioned subsystems. 

\subsection{Ground level}
The core of the ground level consists of a local Decisional Support System (DSS), embedded on a special-purpose processor installed on the \textsc{Heimdall}'s lampposts. We defined such a DSS as the \emph{Lamppost Local Unit} (LLU). It provides an AI module that acquires, in real-time through the on-board cameras, high-resolution images and videos involving the current traffic situation in the observed area. Then, the acquired data are automatically augmented with vehicles speeds and spatial correlations. 

\begin{figure}[htp!]
\centering
\includegraphics[width=0.78\linewidth]{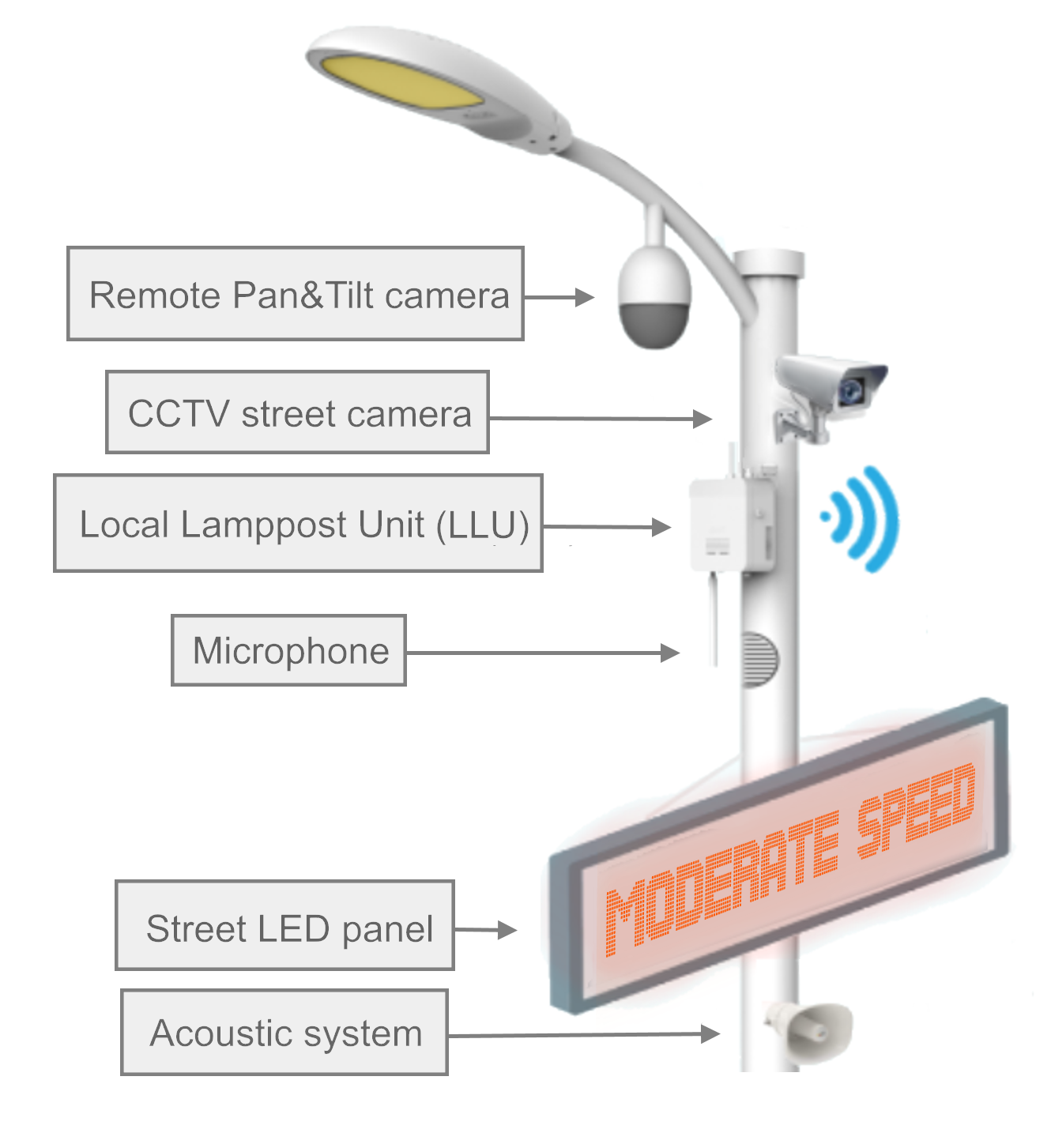}
\caption{Hardware scheme of the smart lamppost.}
\label{fig:lamppost}
\end{figure}

To this purpose, the module on the LLU leverages on a \emph{YOLO (You Only Look Once)} system  \cite{redmon2016you}, a pretrained SSD (Single Shot Detector) based on a deep Convolutional Neural Network (CNN) that, by processing a video frame, returns information on the position and dimensions of any object inside the frame, denoting it with a bounding box. Hence, a second software component, consisting of a Faster R-CNN (detailed in Section \ref{frcnn}) uses the output of YOLO to identify the presence of anomalies. In the event an anomaly is detected, the LLU reports its inferred typology and associated \emph{criticality index} $\phi$ (with $\phi \in [0, N]$), together with the area-specific metadata (audio/video recordings, MSD, \emph{etc.}), to the central unit that constitutes the Territorial level. 

The lamppost is also equipped with sound and light signalling devices which, depending on the severity of the anomaly detected, may be activated in order to invite drivers to moderate their speed, or to signal the accident, as well as receive remote commands from the human supervisor.
Finally, we should note that the complete software that powers the LLU, including the AI module that executes the deep neural networks, runs on a single integrated hardware board, embedded in the lamppost. In our configuration, this board is represented by an NVIDIA \emph{Jetson AGX Xavier}\footnote{\url{https://nvidia.com/en-gb/autonomous-machines/embedded-systems/jetson-agx-xavier/}}, a compact and low-power system-on-module (SoM), which features a Volta GPU, two Deep Learning Accelerators (DLAs), an octa-core Carmel CPU and 32GB of memory, provided with a pre-installed Linux distribution and a dedicated developer kit. The full hardware scheme of the smart lamppost is summarized in Figure \ref{fig:lamppost}.

\subsection{Territorial level}
The territorial level represents a higher level of \textsc{Heimdall}'s architecture: specifically, its core module is mainly composed by the \emph{Territorial Control Unit} (TCU). With respect to the Ground level devices, this unit performs a wider range analysis, since: \textit{i)} it gathers the local information and criticality index coming from the single LLUs, with the aim of validating and managing the local alerts, and \textit{ii)} it is in charge of predicting possible risk situations which could affect the traffic circulation, in order to prevent accidents.

To achieve this goal, the TCU collects data coming from the LLUs installed on smart lampposts, combining them with the information retrieved from external sources (e.g., weather services, public utility information, civil protection alerts). By geo-locating such information, the TCU employs an additional level of AI, primarily based on heuristics and control thresholds, where each criticality index $\phi_i$, sent by the $i$-th LLU, is reassessed on the basis of a \emph{global risk index} $\lambda$ (with $\lambda \in [0, M]$). In formulae:

$$ \varphi_i = \max \{\left \lceil{\phi_i + \alpha \lambda}\right \rceil, N\} $$

where $\varphi_i$ is the \emph{reassessed criticality index} (associated to the local area monitored by the smart lamppost $i$), while $\alpha \in [0, 1]$ is the parameters for weighting the impact of the global index, tuned according to the specific properties of the considered territory. It should also be noted that this index is rounded up, and can never exceed $N$, which represents the maximum level of criticality expected for a certain area.

Given $\varphi$ values, the TCU can therefore operate in several ways. First of all, it can automatically propagate alert signals between proximal smart lamppost (e.g. the active acoustic and light warnings). Secondly, it can activate a \emph{preventive warning}, if it detects the possibility of imminent critical situations, such as rain risk.

\begin{figure}[htp!]
\centering
\includegraphics[width=0.98\linewidth]{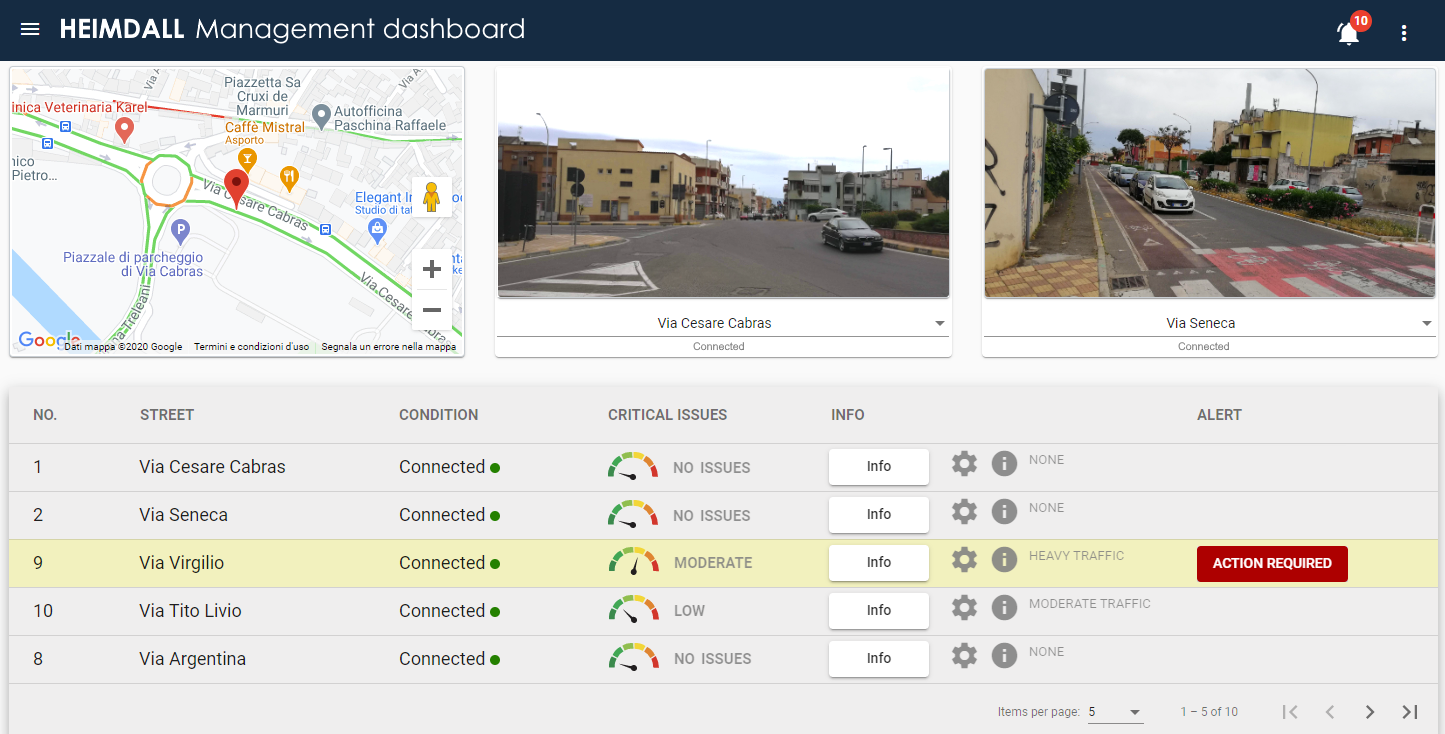}
\caption{The management dashboard of \textsc{Heimdall}.}
\label{fig:dashboard}
\end{figure}

However, although the TCU is set up to handle a large set of anomalies in a fully automatic way, the system is still set up for supervision by a human operator. Through a responsive \emph{management dashboard} (shown in Figure \ref{fig:dashboard}), both accessible from desktop and mobile devices, the operator is able to filter false positives/negatives, thus improving the accuracy of the different AI layers, as well as managing an intelligent queue of notifications related to active alerts (sorted by priority based on timestamp and criticality index), to deactivate or propagate further.

\subsection{Modelling level}
The last tier of \textsc{Heimdall} is the modelling level, through a dedicated Training Module. Its main goal is the definition and training of the models involved in the prediction and classification operations which rule the AI modules of the LLUs and TCU. This subsystem runs on an \emph{ad-hoc} cloud service, and it is not required to work in real-time with the first two tiers. The activities performed by this level can be summarized in four distinct but still interconnected phases:
\begin{itemize}
    \item \textbf{Phase 1}: collecting training data, merged with additional features from outer sources (e.g., weather forecasting databases, historical series providers, and/or further local traffic monitoring systems);
    \item \textbf{Phase 2}: training the involved networks;
    \item \textbf{Phase 3}: validating the models and tuning of the hyper-parameters; testing of the tuned networks (repeat Phase 2 when necessary);
    \item \textbf{Phase 4}: deploying of trained models on the destination units.
\end{itemize}

\section{Anomalies Detection}
\label{sec:anomalies}

The main issues faced by means of the \textsc{Heimdall} system are related to traffic monitoring tasks, with special attention to the detection of anomalous events in Smart Cities and urban scenarios. This class of problems can be handled in two different ways:
\begin{itemize}
    \item the first one requires to build the model by learning from the scene which events can be considered as \emph{normal}, in order to detect those which deviate from the regular behavior \cite{bergmann2019mvtec};
    \item the second one requires to address a supervised object detection problem: in this case, the model is trained over the anomalous events, in order to generalise on the new videos.
\end{itemize}
We observe that there exist several pros and cons in relation to both the approaches: the first one could potentially be more accurate, but it needs a very considerable training phase, in order to let the system discriminate which scene contains an anomaly. Moreover, in this context, if a particular anomaly occurs with regularity, it can be missed by the model.
The second approach, which is currently a preliminary proposal, is the one adopted for the \textsc{Heimdall} infrastructure: a set of anomalies is listed \emph{a-priori}; then, all items in this set are labelled and passed as a training set to a Convolutional Neural Network (CNN), in charge of learning the pattern related to each type of anomaly. The adopted CNN architecture is a Faster R-CNN, which is discussed in details in the remainder of this section.

\subsection{Dataset and annotated anomalies}
In order to achieve a reliable training of the network, we have taken into consideration some preliminary factors: (i) the population of the dataset; indeed, in order to achieve a proper anomalies detection in urban scenarios, a variegated dataset needs to be used and labelled too. At the state-of-the-art, some datasets exist, but either proper annotations are missing or the camera view is not suitable for video surveillance systems \cite{yao2020and}; (ii)
the kind of anomalies to be detected; in most scenarios, only anomalies leading to extreme events are annotated, e.g. accidents or pedestrians hitting. However, since the anomalies we aim to detect cover wider aspects of the urban education, we decided to proceed with an \emph{ad-hoc} acquired dataset, so to include a large number of specific events.

\begin{figure}[htp!]
\centering
\includegraphics[width=0.9\linewidth]{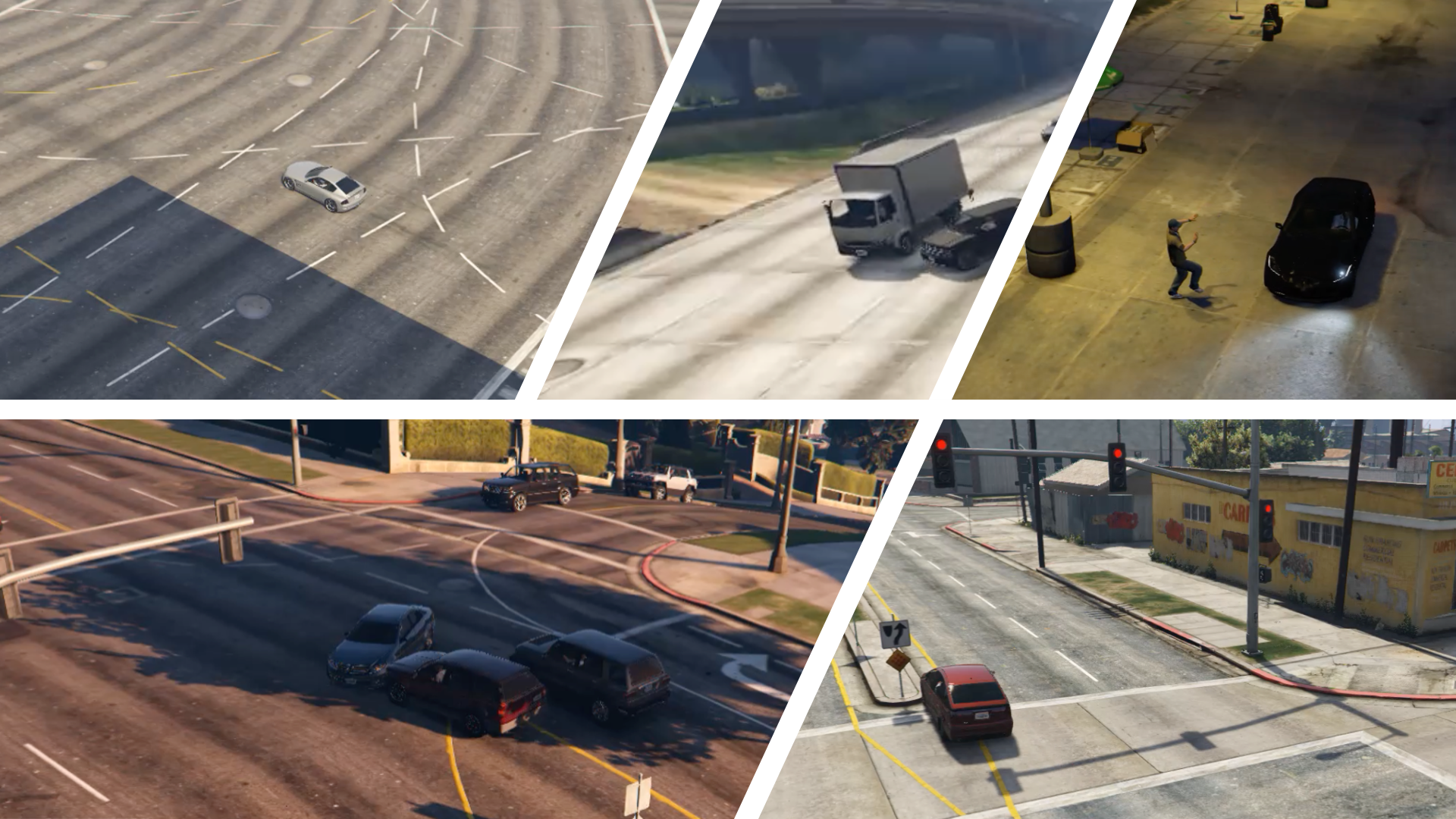}
\caption{Five kinds of annotated anomalies of the collected dataset. From the top-right, clockwise: \emph{illegally parked vehicle}, \emph{risky overtaking}, \emph{vehicle on pedestrian area}, \emph{red light violation}, \emph{vehicle-vehicle collision}.}
\label{fig:anomalies}
\end{figure}

Hence, for the training phase, we built and annotated a large set of videos (currently, they are $\sim500$, but the number is still increasing), featuring several types of events, which can be considered anomalous or not depending on the kind of scenario application. Such videos are synthetically generated, and acquired from the \textit{Grand Theft Auto V} videogame from Rockstar Games. This solution provided a good trade-off between the realism of the simulations and the configuration of the acquisition, along with the simplicity and the velocity of reproducibility of different, even rare, events. We also observe that such technique has already shown to be very effective in the literature, as in \cite{fabbri2018learning}, where the authors created the JTA (Joint Track Auto) dataset to estimate subject poses. In Figure \ref{fig:anomalies}, some examples of the annotated anomalies are shown.

\subsection{Faster R-CNN} \label{frcnn}
The Training module of the modelling level, detailed in the previous section, involves the training of a Faster R-CNN \cite{ren2015faster}, with a backbone consisting of a VGG-16 net. This kind of network is a Region Proposal Network (RPN), since it involves a two step process for detecting objects within a scene:
\begin{itemize}
    \item the first step is the region proposal one, in which a set of regions within the image is firstly selected for subsequent classification;
    \item the second step is in charge of classifying the single regions, in order to verify whether an object of interest is present.
\end{itemize}

\begin{figure}[htp!]
\centering
\includegraphics[width=0.95\linewidth]{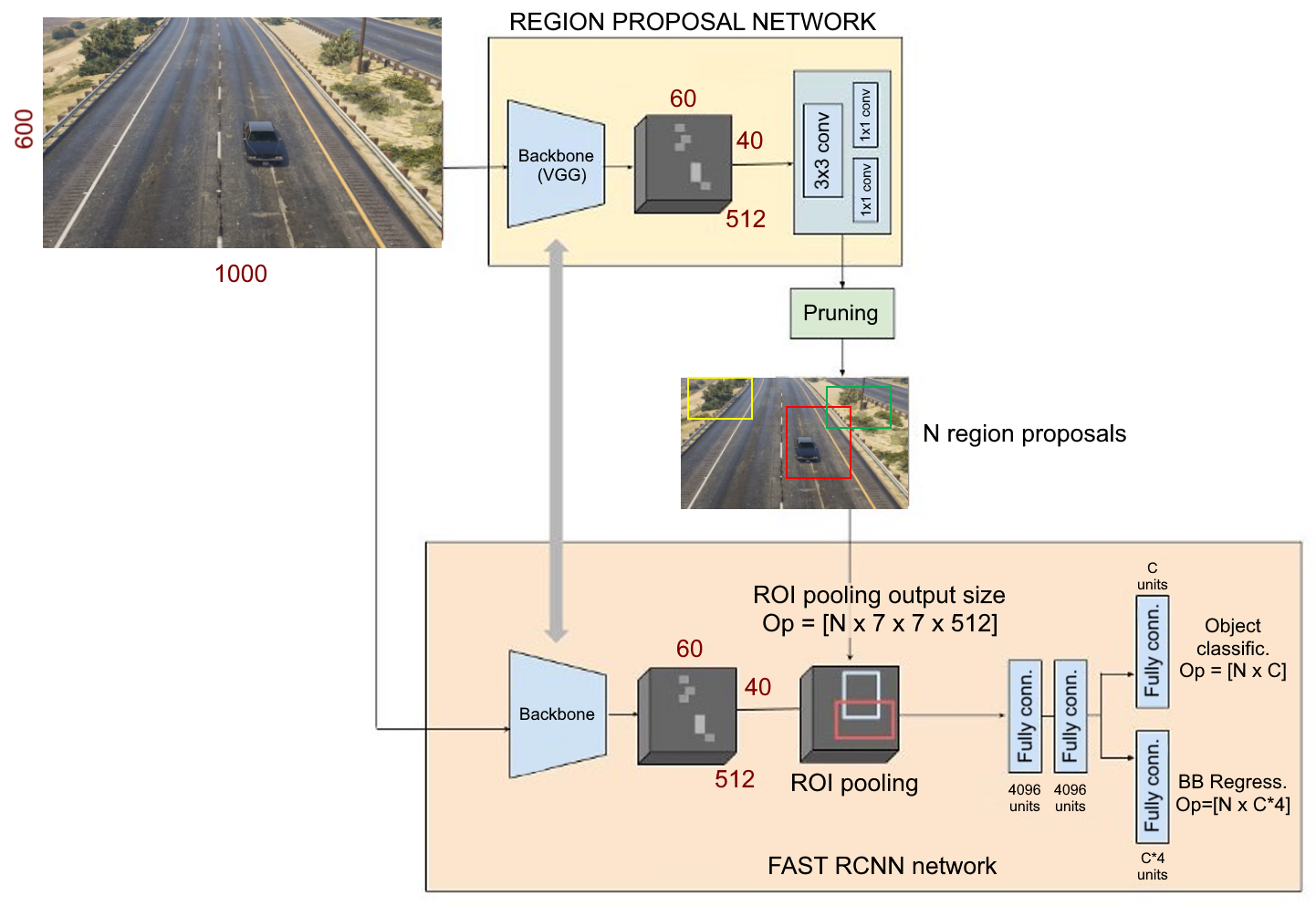}
\caption{The Faster R-CNN used in the Training Module for classifying events.}
\label{fig:cnn}
\end{figure}

The architecture of such a network is depicted in Figure \ref{fig:cnn}, where the Region Proposal Network and the Faster R-CNN network respectively achieve the aforementioned steps. Please refer to the work in \cite{ren2015faster} for Faster R-CNN implementation and architecture details.  

\begin{figure}[htp!]
\vspace{0.2em}
\centering
\includegraphics[width=0.7\linewidth]{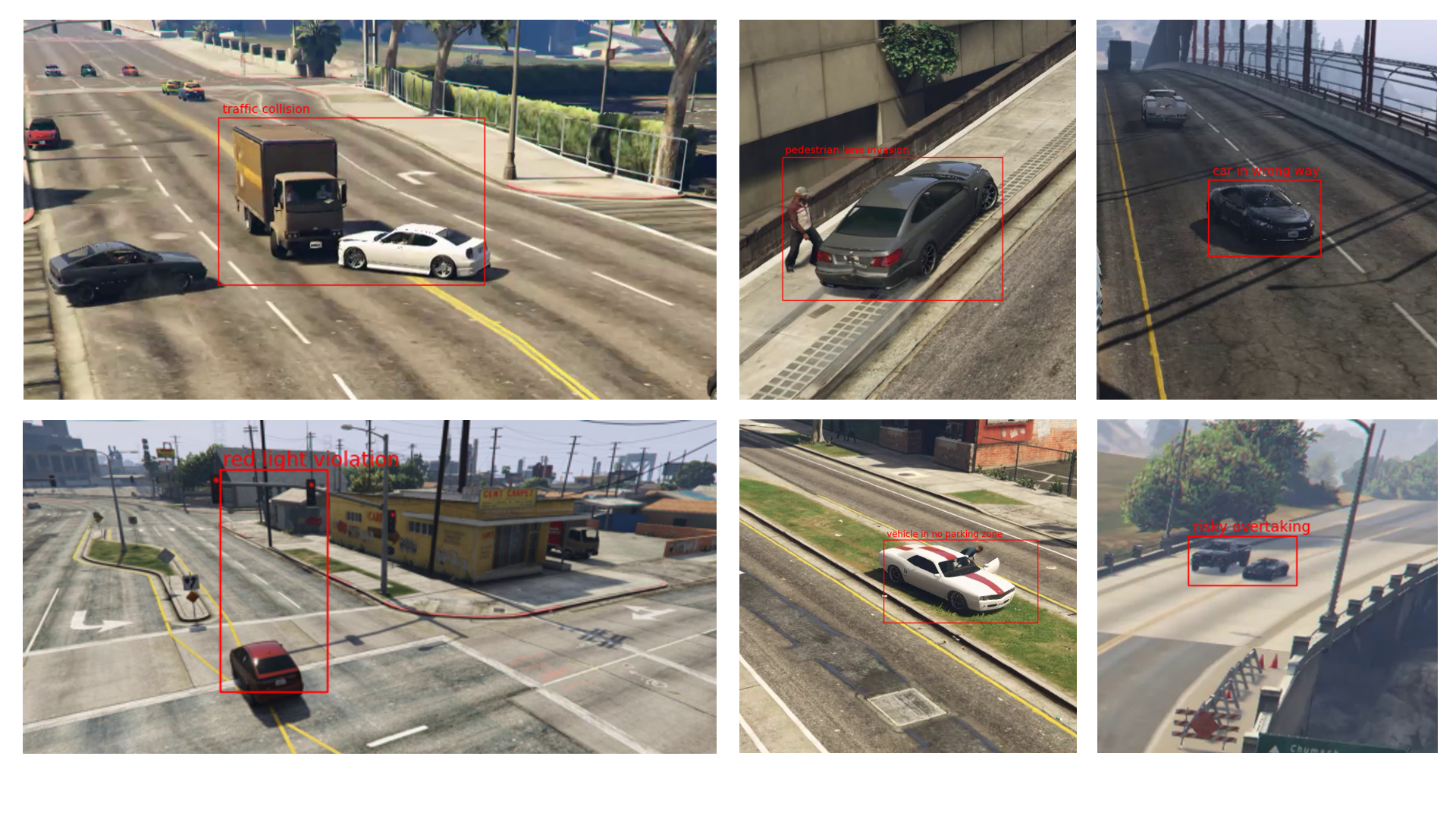}
\caption{Some examples of correct anomaly detection, when the trained Faster R-CNN is executed on synthetic test videos. From the top-right, clockwise: \textit{traffic collision}, \textit{vehicle on pedestrian area}, \textit{car in wrong way}, \textit{risky overtaking}, \textit{car in no parking area} and \textit{red light violation}.}
\label{fig:gta_res}
\end{figure}

\begin{figure}[htp!]
\centering
\includegraphics[width=0.7\linewidth]{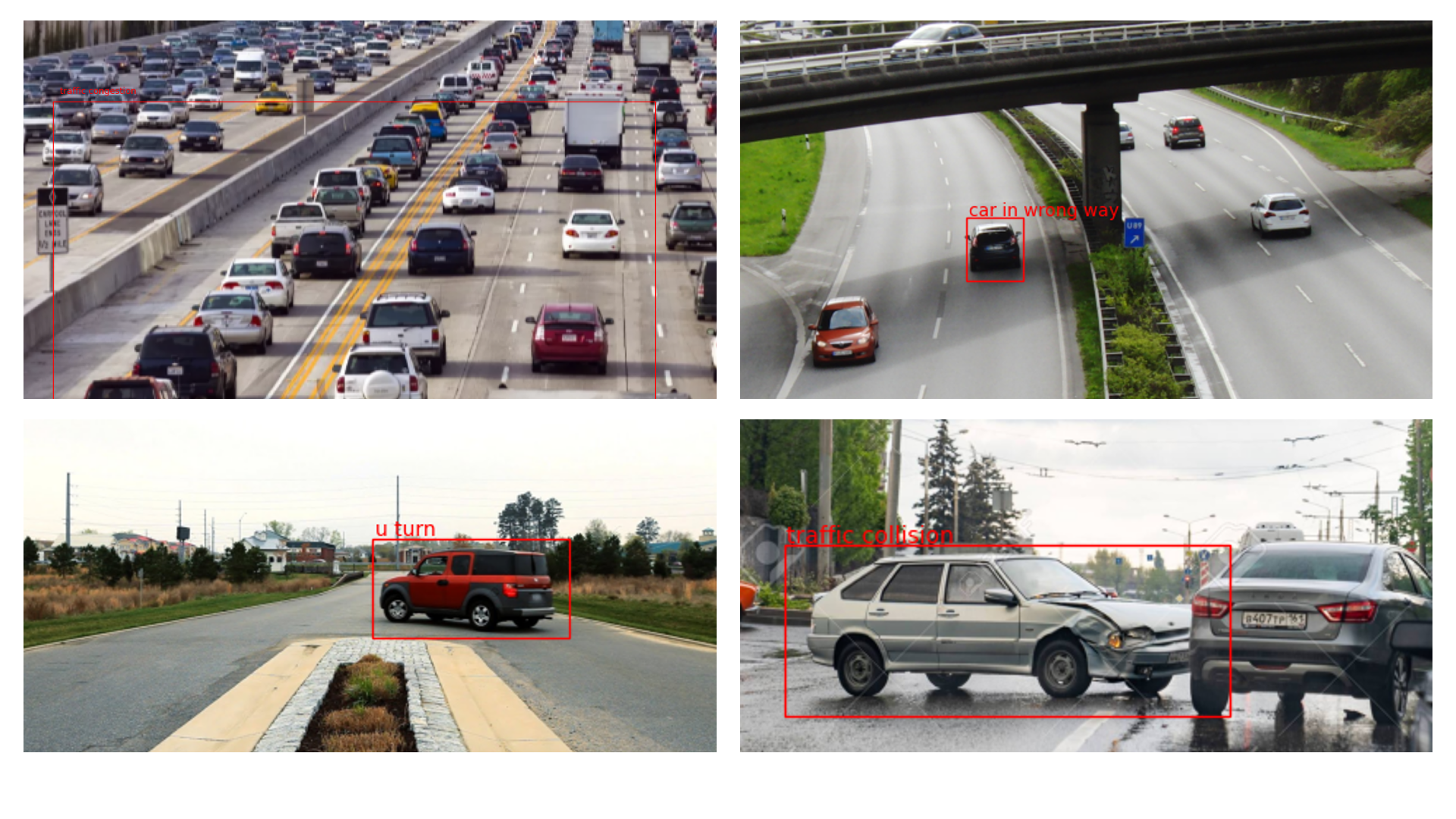}
\caption{Some examples of correct anomaly detection, when the trained Faster R-CNN is executed on real-world videos. From the top-right, clockwise: \textit{traffic congestion}, \textit{car in wrong way}, \textit{traffic collision} and \textit{risky u-turn}.}
\label{fig:real_res}
\end{figure}

In order to fasten the deploy of the network, we exploited the \texttt{detecto} framework, to allow a quick training phase configuration, thus producing a trained network. In order to properly use the framework, the images have been resized to $600 X 1000$. The dataset has been labelled by means of the tool \texttt{LabelImg}, which facilitates the selection and annotation of the regions within the frames of a video. The preliminary qualitative results obtained, shown in Figure \ref{fig:gta_res}, reveal a twofold interesting insight:
\begin{itemize}
    \item on synthetic videos, the network results very accurate, especially on \emph{static} anomalies (which do not imply a time component); but, nonetheless, some dynamic anomalies are still well detected (e.g., \emph{overtaking} or \emph{U-turns}). Due to the adopted approach, all the anomalies related with vehicle speed are not currently detected, since we adopted no method for relating consecutive frames at this stage;
    \item even when executed on real scenario videos, the Faster R-CNN is able to detect most of the anomalies, especially when the camera view is similar to that of dataset samples and no major occlusions are present, even under dense traffic conditions. Some examples are shown in Figure \ref{fig:real_res}.
\end{itemize}

\vspace{-0.5em}

\section{Discussion, Conclusions and Limitations}
\label{sec:conclusions}

We have presented \textsc{Heimdall}, an AI-based video surveillance infrastructure composed by three tier, respectively in charge of \textit{i)} sensing the urban environment, \textit{ii)} coordinating the communications and \textit{iii)} training the AI models involved. Our proposed infrastructure employs a YOLO system, at the ground level, for tracking vehicles and pedestrians, and exploits a preliminary approach based on a Faster R-CNN, for the detection of potential anomalies in the scene.  Specifically, the learning of such a CNN is in charge of the modelling level, so to separate the computing burden needed for the training phase, from the actual deployment step, which takes place on the Jetson AGX Xavier card embedded in the lamppost. 

The anomalies detection module has shown interesting insights from a qualitative point of view: even though it was mainly trained on synthetic videos (generated by means of the Rockstar editing tool over the \textit{GTA-V} videogame), it has proven to be capable of providing good results also in the real-world scenario. However, it still presents some limitations, mainly due to two aspects, which have become clearer by following the downstream analysis on the working tools: (i) some anomalies which implies dynamic analysis are cumbersome to detect, since, natively, the Faster R-CNN does not take into consideration the relationships between the objects in consecutive frames; (ii) to improve the accuracy of the detection, particularly in the presence of widely differing scenes, it could be useful to exploit a domain adaptation approach\cite{kashiani2020online} to keep unaltered the training phase, while modifying internally the learnt filters and adapting them to the new scenario. 
These elements represent significant milestones for further improvements, and they are currently planned for a future release of the discussed infrastructure.

%\section*{Acknowledgment}

%\textcolor{red}{DOBBIAMO CHIEDERE A SALVATORE}

\bibliographystyle{IEEETran}
\bibliography{refs}

\end{document}